\documentclass[aps,prd,twocolumn,floatfix,nofootinbib,superscriptaddress]{revtex4-2}
\usepackage[colorlinks,linkcolor=blue,anchorcolor=blue,citecolor=blue,urlcolor=blue,breaklinks=true]{hyperref}
\usepackage[libertine]{newtxmath}
\usepackage{graphicx}
\usepackage{slashed}
\usepackage{amsfonts}
\usepackage{amsmath}
\usepackage{blindtext}
\usepackage{microtype}

\newcommand{\vect}[1]{\boldsymbol{#1}}
\newcommand{\ph}[1]{\phantom{#1}}
\newcommand{\sh}[1]{\slashed{#1}}

\graphicspath{{.}{figures/}}

\def\hs{\hspace}

\def\lf{\left}
\def\rg{\right}

\begin{document}

\title{Heavy flavor-asymmetric pseudoscalar mesons on the light front}

\author{Chao Shi}
\email[]{cshi@nuaa.edu.cn}
\affiliation{Department of Nuclear Science and Technology, Nanjing University of Aeronautics and Astronautics, Nanjing 210016, China}

\author{Pengfei Liu}
\affiliation{Department of Nuclear Science and Technology, Nanjing University of Aeronautics and Astronautics, Nanjing 210016, China}

\author{Yi-Lun Du}
\email[]{yilun.du@iat.cn}
\affiliation{Shandong Institute of Advanced Technology, Jinan 250100, China}

\author{Wenbao Jia}
\affiliation{Department of Nuclear Science and Technology, Nanjing University of Aeronautics and Astronautics, Nanjing 210016, China}

\begin{abstract}
We extract the leading Fock-state light front wave functions (LF-LFWFs) of  heavy flavor-asymmetric pseudoscalar mesons $D$, $B$ and $B_c$ from their Bethe-Salpeter wave functions based on Dyson-Schwinger equations approach, and study their leading twist parton distribution amplitudes, generalized parton distribution functions and transverse momentum dependent parton distributions. The spatial distributions of the quark and antiquark on the transverse plane are given, along with their charge and energy distributions on the light front. We find that in the considered mesons, the heavier quarks carry most longitudinal momentum fraction and yield narrow $x$-distributions, while the lighter quarks play an active role in shaping the transverse distributions within both spatial and momentum space, exhibiting a duality embodying characteristics from both light mesons and heavy quarkonium.
\end{abstract}
\maketitle

%===============================================================================
%===============================================================================
\section{INTRODUCTION\label{intro}}
  
The heavy-light pseudoscalar mesons, such as $D$ and $B$ mesons, are intriguing QCD bound states with significant flavor asymmetry. They provide a rich area of study for CP violation effects and have thus garnered substantial experimental and theoretical interest \cite{BaBar:2001pki, Kobayashi:1973fv, Belle:2001zzw, Lunghi:2002ju, Bosch:2003fc,Beneke:2018wjp,LHCb:2019hro}. On the other hand, while they are often likened to the hydrogen atom in QED due to the mass difference between their internal constituents, these mesons exhibit significant non-perturbative QCD dynamics and relativistic effects that cannot be ignored. Studying the partonic structure of the heavy-light mesons is thus important for revealing their role in high energy processes. 

The light-front wave functions (LFWFs) are crucial for characterizing hadron's partonic structure. They are the basic wave functions of hadrons formulated with light-front coordinate, and offer intuitive interpretations similar to Schr$\ddot{\rm o}$dinger wave functions in non-relativistic quantum mechanics. Through them, various parton distributions that are physically meaningful can be calculated, including the parton distribution amplitude (DA), which serves as an important nonperturbative element in the factorization of $D$ and $B$ meson exclusive decays \cite{Beneke:2018wjp,Braun:2012kp,Beneke:2000ry,Bosch:2001gv,Braun:2003wx,Han:2024min}, the generalized parton distributions (GPD) and transverse momentum distributions (TMD) that shed light on the three-dimensional (3D) parton distribution inside hadrons \cite{Ji:1996nm,Radyushkin:1997ki,Burkardt:2002hr,Angeles-Martinez:2015sea,Boussarie:2023izj}.

The LFWFs of heavy flavor-asymmetric mesons have been explored by various works. For the $B$ mesons, where heavy quark effective theory (HQET) is usually applied, the interest mostly focused on LFWFs defined with effective $b$-quark field. $B$ meson WFs with Wandzura-Wilczek approximation \cite{Wandzura:1977qf} and beyond are usually considered \cite{Wu:2013lga,Huang:2005kk}. For $D$ meson and lighter quarks, LFWFs defined with normal quark field are considered. Modeling and Melosh rotation techniques are usually employed in generating the nonperturbative dynamics and spin configurations \cite{Melosh:1974cu}.

On the other hand, the flavor asymmetric mesons' LFWFs have been studied by the basis light front quantization (BLFQ) method, including the $D$, $B$ and $B_c$ mesons, as well as their excitations. By solving a revised holographic QCD Hamiltonian devised for heavy quarks, the meson spectrum, decay constants and LFWFs are obtained \cite{Tang:2019gvn,Tang:2018myz}. Meanwhile, the spectrum, decay constant \cite{Chen:2019otg,Serna:2017nlr,Gomez-Rocha:2016cji,El-Bennich:2016qmb,Qin:2019oar}, parton distribution amplitudes \cite{Serna:2020txe,Binosi:2018rht} and electromagnetic form factors \cite{Xu:2024fun} of flavor asymmetric mesons had been studied with Dyson-Schwinger and Bethe-Salpeter equations (DS-BSEs) approach,  while the investigation of LFWFs remains unexplored. This paper will deliver DS-BSEs prediction on leading Fock-state light front wave functions (LF-LFWFs) of heavy flavor-asymmetric pseudoscalar mesons, including $D$, $B$ and $B_c$ mesons, and study their multi-dimensional parton image.

The method we employ to extract LF-LFWFs in this work is by projecting the covariant Bethe-Salpeter wave functions in ordinary space-time onto the light front, namely, by setting the light front time $x^+\equiv \frac{1}{\sqrt{2}}(x^0+x^3)=0$ in Bethe-Salpeter wave function. It does not take the standard light-front quantization procedure, yet also has a long history in practice \cite{tHooft:1974pnl,Liu:1992dg,Lepage:1980fj,Burkardt:2002uc,Mezrag:2016hnp}, including a very recent implementation in lattice QCD \cite{LatticeParton:2023xdl}. Based on this method and within DS-BSEs approach, we have studied the LF-LFWFs of various light and/or heavy mesons, such as pion, $\rho$, $\eta_c$ and $J/\psi$ mesons \cite{Shi:2018zqd,Shi:2021nvg, Shi:2021taf,Shi:2022erw, Shi:2023oll,Shi:2023jyk}. The extension of investigation to mesons with large flavor asymmetry is thus straightforward. Physically, it is interesting to see how the flavor asymmetry affects the LF-LFWFs. In particular, in $D$ and $B$ mesons, the dynamical chiral symmetry breaking (DCSB) strongly reshapes the light quarks, yet it is suppressed in the heavy quarks where Higgs mechanism dominates. The heavy-light mesons thus host two important mechanisms simultaneously, offering a good ground to see their interplay and manifestation in one bounded system.

This paper is organized as follows: In Sec.~\ref{sec:BSE} we introduce the DS-BSEs formalism and the treatment of heavy-light mesons. In Sec.~\ref{sec:bs2lf}, the definition of LF-LFWFs and their extraction from the covariant BS wave functions are given. In Sec.~\ref{sec:3d}, we utilize the LF-LFWFs to study the 3D parton structure of flavor-asymmetric mesons with GPDs (at zero skewness) and TMDs. Finally we conclude in Sec.~\ref{sec:con}

\section{Bethe-Salpeter wave functions of flavor asymmetric pseudoscalar mesons\label{sec:BSE}}
\label{sec:2}
We employ the Bethe-Salpeter wave functions of heavy-light meson referring to the treatment in \cite{Xu:2024fun}, which were obtained by numerically solving full quark propagator's Dyson-Schwinger equation and meson's Bethe-Salpeter equation. In the following we recapitulate the method. The quark propagator's Dyson-Schwinger equation under the rainbow ladder (RL) truncation reads
\begin{equation}
  S^{-1}(p)=Z_2i \gamma \cdot p + Z_4 m + Z_{1} \int_q^{\Lambda} g^2 D_{\alpha \beta}(p-q) \frac{\lambda^a}{2} \gamma_\alpha S(q) \frac{\lambda^a}{2} \gamma_\beta,
  \label{eq.DSE}
\end{equation} 
with $S^{-1}(p)$ the full quark propagator. The $m$ is the current quark mass and $Z_{1,2,4}$ are the renormalization constants. A mass-independent momentum-subtraction renormalization scheme \cite{Chang:2008ec} is employed, and  a translationally-invariant regularization of the four-dimensional integral  $\int_q^{\Lambda}$ is used. For the dressed-gluon propagator $D_{\mu\nu}(k)$, the Qin-Chang model \cite{Qin:2011xq} is used, which takes the replacement
\begin{equation}
  Z_1 g^2 D_{\mu \nu}(k)=Z_2^2 \mathcal{G}\left(k^2\right) (\delta_{\mu \nu}-{k_\mu k_\nu}/k^2)\equiv {\cal D}_{\textrm{eff}}(k),
  \label{eq.gluon}
\end{equation}
where
\begin{align}
\frac{\mathcal{G}(k^2)}{k^2}&=\mathcal{G}^{\text{IR}}(k^2)+\frac{8\pi^2\gamma_{m}\mathcal{F}(k^2)}{\ln[\tau+(1+k^2/\Lambda^2_{\text{QCD}})^2]}, \\
\mathcal{G}^{\text{IR}}(k^2)&=D\frac{8\pi^2}{\omega^4}e^{-k^2/\omega^2},
\end{align}
with $\mathcal{F}(k^2)=\{1-\exp[(-k^2/(4m_t^2)]\}/k^2$, $m_t=0.5$\,GeV, $\tau=e^2-1$, $\Lambda_{\text{QCD}}=0.234$\,GeV, $\gamma_{m}=12/25$ \cite{Xu:2021mju}.  Following \cite{Xu:2024fun}, we use $(D\omega)_{u/d}=(0.82\ \textrm{GeV})^3$, $(D\omega)_c=(0.66\ \text{GeV})^3$ ,$(D\omega)_b=(0.48\ \text{GeV})^3$, with current quark masses  $m_{u/d}=0.0033$ GeV, $m_{c}=0.854$ GeV and $m_b=3.682$ GeV renormalized at $19$ GeV. They are fine tuned to reproduce the physical meson spectrum and electroweak decay constants, which can be found in Table.~2 of \cite{Xu:2024fun}. The fully dressed quark propagator can be decomposed as 
\begin{equation}\label{eq:S}
  S^{-1}(p)= i\gamma \cdot pA(p^2) + B(p^2).
\end{equation}
The scalar functions $A(p^2)$ and $B(p^2)$ can be obtained by numerically solving the quark gap Eq.~(\ref{eq.DSE}).

On the other hand, the meson's Bethe-Salpeter amplitude, $\Gamma^{f\bar{h}}_M\left(k; P\right)$ is obtained by solving the BSE, i.e.,
\begin{equation}
  \Gamma^{f\bar{h}}_M\left(k; P\right) = \int_{q}^{\Lambda} K^{f\bar{h}}(q,k;P)  S^{f}\left(q_\eta\right) \Gamma^{f\bar{h}}_M\left(q; P\right) S^{h}\left(q_{\bar{\eta}}\right),
    \label{eq:BSE}
\end{equation}
where $f$ and $h$ denote the flavor of (anti-)quark, $k_{\eta}= k + \eta P$, $k_{\bar{\eta}}= k - (1-\eta) P$ with the momentum partitioning parameter $\eta \in [0,1]$. Note that physical observables do not depend on $\eta$. This gives us the freedom to choose appropriate value for $\eta$ in numerical calculation, which is critical in solving the flavor-asymmetric meson BSE by avoiding singularities in dressed quark propagators \footnote{See, e.g., \cite{Rojas:2014aka,Serna:2020txe,Xu:2024fun} for more details about singularity behavior of dressed quark propagators on the complex plane.}. It is important to note that the homogeneous BS equation allows an arbitrary multiplicative factor for BSA. In physics, the BSA \emph{must} satisfy canonical normalization condition. Its physical meaning is that the valence quark number within a meson should add up to one \cite{Nakanishi:1969ph}. Under the RL truncation, the normalization condition reads
\begin{align}
2 P_\mu&=\int_k^\Lambda \left\{{\rm Tr}\left[\bar{\Gamma}_M^{f\bar{h}}(k;-P)\frac{\partial S_f(k_\eta)}{\partial P_\mu} \Gamma_H^{f\bar{h}}(k;P)S_h(k_{\bar{\eta}})\right] \right. \nonumber \\
&\hspace{10mm}+\left. {\rm Tr}\left[\bar{\Gamma}_M^{f\bar{h}}(k;-P)S_f(k_\eta)\Gamma^{f\bar{h}}_M(k;P)\frac{\partial S_h(k_{\bar{\eta}})}{\partial P_\mu} \right]\right \},
\end{align}
with $\bar{\Gamma}_M(k,-P)^T=C^{-1}\Gamma_M(-k,-P)C$ and $C=\gamma_2 \gamma_4$. 

The specification of the interaction kernel $K^{f\bar{h}}(q,k;P)$ in Eq.~(\ref{eq:BSE}) is an open question that is not fully resolved. The key issue is how to find a $K^{f\bar{h}}(q,k;P)$ that rigorously preserves the axial-vector Ward Takahashi identity, as required by QCD. Various schemes were proposed \cite{Qin:2019oar,Serna:2017nlr,Chen:2019otg} yet the problem remains unsolved in a rigorous sense. Here we take the treatment from \cite{Xu:2024fun} which linearly combines the Rainbow-Ladder interaction kernel of flavor $f$ and flavor $h$, i.e.,
\begin{equation}
\label{eq:kernel.wRL}
  K^{f\bar{h}}(q,k;P) = \alpha_{f\bar{h}} K_{\text{RL}}^{f\bar{f}}(q,k;P) + (1-\alpha_{f\bar{h}}) K_{\text{RL}}^{h\bar{h}}(q,k;P),
 \end{equation}
 with 
\begin{equation}
\label{eq:kernel.RL}
  K^{f\bar{f}/h\bar{h}}_{\text{RL}}(q,k;P) = \mathcal{D}^{f/h}_{\text{eff}}\frac{\lambda^a}{2} \gamma_\alpha \otimes \frac{\lambda^a}{2} \gamma_\beta.
\end{equation}
The $\alpha_{f\bar{h}}$ is a model parameter introduced to accommodate the flavor-asymmetric effect in heavy-light systems. They are determined to be $\alpha_{c[\bar{u}/\bar{d}]}=0.689$, $\alpha_{b[\bar{u}/\bar{d}]}=0.751$ and $\alpha_{b\bar{c}}=0.585$, which help produce the mass spectrum and decay constants reported in Table.~2 of \cite{Xu:2024fun}.
The $\Gamma_M^{f\bar{h}}(k;P)$ can be most generally decomposed as \cite{Dai:1993qr}
\begin{align}
\label{eq:gammapara}
\Gamma_M^{f\bar{h}}(k;P) &= \gamma_5 \Big[i E(k;P)+\sh{P} F(k;P)\nonumber \\
&\hs{10mm}
+(k \cdot P)\sh{k}\,G(k;P)+i [\sh{k},\sh{P}]\,H(k;P)\Big]. 
\end{align}
The ${\cal F}=E,F,G$ and $H$ are scalar functions of $k^2, k\cdot P$ and $P^2=-m_H^2$. The $m_H$ is meson's mass and the minus sign is due to Euclidean space we take. Then one can obtain the numerical solution of ${\cal F}(k;P)$'s by solving Eq.~(\ref{eq:BSE}), with input from Eqs.~(\ref{eq.gluon},\ref{eq:S},\ref{eq:kernel.wRL}). A detailed introduction to the calculation and computation techniques can be found in \cite{Blank:2011qk}. To conclude, we now have the fully numerical solutions to the dressed quark propagator $S(p)$ and Bethe-Salpeter amplitude $\Gamma(k;P)$ in Eqs.(\ref{eq:S},\ref{eq:gammapara}). The Bethe-Salpeter wave function $\chi_{f\bar{h}}(k,P)$, which takes the momentum partition $k_\eta$ and $k_{\bar{\eta}}$ for quark and antiquark respectively, can be obtained by $\chi_{f\bar{h}}(k;P) = S_f(k_\eta)\,\Gamma_{M}(k;P)\,S_h(k_{\bar{\eta}})$.

\section{From Bethe-Salpeter wave functions to light front wave functions}\label{sec:bs2lf}
The light-front wave functions of a hadron are defined as the coefficient amplitudes associated with Fock-state expansion, which encode nonperturbative dynamics within a hadron on the light-front. For instance, meson $M$ with valence quark $f$ and anti-quark $\bar{h}$ at the leading Fock-state reads \cite{Brodsky:1997de}
\begin{align}\label{eq:LFWF1}
|M\rangle &= \sum_{\lambda_1,\lambda_2}\int \frac{d^2 \vect{k}_T}{(2\pi)^3}\,\frac{dx}{2\sqrt{x\bar{x}}}\, \frac{\delta_{ij}}{\sqrt{3}} \nonumber \\
&\hspace{10mm} \Phi_{\lambda_1,\lambda_2}(x,\vect{k}_T)\, b^\dagger_{f,\lambda_1,i}(x,\vect{k}_T)\, d_{h,\lambda_2,j}^\dagger(\bar{x},\bar{\vect{k}}_T)|0\rangle,
\end{align}
where the $b^\dagger$ and $d^\dagger$ are the creation operators for quark and antiquark respectively, and the $\Phi_{\lambda_1,\lambda_2}(x,\vect{k}_T)$ are the LFWFs. In a frame where the hadron has no transverse momentum, the quark with flavor $f$ carries transverse momentum $\vect{k}_T$ and longitudinal momentum $k^+=x P^+$. By momentum conservation, the antiquark carries $\bar{x}=1-x$ and  $\bar{\vect{k}}_T=-\vect{k}_T$. The $\lambda$ is (anti)quark helicity and runs through $\uparrow,\downarrow$. The $i,j$ are the color indicies.  
As pointed in \cite{Ji:2003yj}, the four $\Phi_{\lambda_1,\lambda_2}(x,\vect{k}_T)$'s, with $\lambda_1,\lambda_2$ running through $\uparrow,\downarrow$ respectively, can be expressed with two independent scalar amplitudes due to symmetry constraints, i.e.,
\begin{align}\label{eq:phiT}
\!\!\!\! \Phi_{\uparrow,\downarrow}(x,\vect{k}_T)&=\psi_0(x,\vect{k}_T^2),&\Phi_{\downarrow,\uparrow}(x,\vect{k}_T)=-\psi_0(x,\vect{k}_T^2), \notag \\
\!\!\!\!  \Phi_{\uparrow,\uparrow}(x,\vect{k}_T)&=k_T^{(-)}\psi_1(x,\vect{k}_T^2), &\Phi_{\downarrow,\downarrow}(x,\vect{k}_T)=k_T^{(+)}\psi_1(x,\vect{k}_T^2),
\end{align}
where $k_T^{(\pm)}=k^1\pm ik^2$. The subscripts $0$ and $1$ of $\psi(x,\vect{k}_T^2)$ indicate the absolute value of orbital angular momentum (OAM) of the system projected in $z-$direction. For instance, the $\Phi_{\uparrow,\uparrow}$ contains quark and antiquark both polarized in $z-$direction, so by angular momentum conservation, the OAM in $z-$direction is $l_z=-1$, which can also be read from the factor $k_T^{(-)}$ associated. 

The $\psi_0$ and $\psi_1$ can be obtained by projecting the Bethe-Salpeter wave functions on to the light front \cite{Mezrag:2016hnp,Xu:2018eii,Shi:2018zqd}
\begin{align}
\label{eq:psi0}
\psi_0(x,\vect{k}_T^2)&= \ph{-}\sqrt{3}\,i\!\int \frac{dk_\eta^+dk_\eta^-}{2\,\pi} \nonumber \\
& \hspace{11mm} 
\textrm{Tr}_D\!\left[ \gamma^+ \gamma_5 \chi_{f\bar{h}}(k,P)\right]  \delta\left(x\,P^+ -k_\eta^+\right),  \\
\label{eq:psi1}
\psi_1(x,\vect{k}_T^2)&= -\sqrt{3}\,i\!\int \frac{dk_\eta^+dk_\eta^-}{2\,\pi}\,  \frac{1}{\vect{k}_T^2} \nonumber \\ 
& \hspace{11mm} 
\textrm{Tr}_D\left[ i\sigma^{+ i}\, \vect{k}_{Ti}\, \gamma_5\, \chi_{f\bar{h}}(k,P) \right] 
 \delta\left(x\,P^+ -k_\eta^+\right),   
\end{align}
where $\sigma^{+ i}=\frac{i}{2}[\gamma^+,\gamma^i]$ and the plus component of a four vector $v$ takes the notation $v^+=(v^0+v^3)/\sqrt{2}$. The trace is taken in Dirac space.  Note that the $\psi_0$ and $\psi_1$ has no dependence on momentum partition $\eta$. This can be seen as $k_\eta=k+\eta P$ and we take a frame where $\vect{P}_T=0$, hence $\eta$ has no influence on $\vect{k}_T$. Meanwhile translational invariance in the momentum integration $\int_{-\infty}^\infty dk^+_\eta dk^-_\eta$ ensures the $\eta-$dependence is totally dropped out.

To obtain the $\psi_0$ and $\psi_1$, we  compute their Mellin moments, and then reconstruct $\psi_0$ and $\psi_1$ by fitting the moments. To be specific, we compute the $(2x-1)$-moments of $\psi_{(i)}(x,\vect{k}_T^2)$ at certain $|\vect{k}_T|$, i.e.,
\begin{align}\label{eq:mom}
\langle (2x-1)^m \rangle_{|\vect{k}_T|}^{(i)}&=\int_0^1 dx (2x-1)^m \psi_{(i)}(x,\vect{k}_T^2),
\end{align}
with $m=0,1,2,..,m_{\textrm{Max}}$. This allows transferring to Euclidean space and the Dirac delta function can be analytically integrated out. It is well-known that high Mellin moments are susceptible to numerical noise. To compute as many Mellin moments as possible—i.e., to achieve an $m$ as large as possible in Eq.~(\ref{eq:mom}), we sample many points when discretizing the scalar functions $E,F,G$ and $H$ of $\chi_{f\bar{h}}(k;P)$ in Eq.~(\ref{eq:gammapara}). The numerical accuracy is checked by doubling the number of sampling points and verifying the numerical stability of the computed moments. With the computing resource we have, $m_{\textrm{Max}}=5$ to $8$ can be achieved, depending on the choice of $\vect{k}_T$ and meson type.

To reconstruct the LFWFs, at every discretized $|\vect{k}_T|$, we fit the  Mellin moments with a unique parameterization form
\begin{align}\label{eq:psiansatz}
h(x)=x^\alpha (1-x)^\beta (c_0+c_1 x+c_2 x^2).
\end{align}
By optimizing the $\chi$, which is defined as
\begin{align}
\chi=\sqrt{\frac{1}{m_{\textrm{Max}}}\sum_{m=0}^{m_\textrm{Max}} \left(\frac{\langle(2x-1)^m\rangle_h-\langle(2x-1)^m\rangle_{\textrm{Cal}}}{\langle(2x-1)^m\rangle_{\textrm{Cal}}}\right)^2}
\end{align}
we can find good fittings with $\chi$ all $\lesssim 1\%$. Note the subscript $h$ refers to Mellin moments calculated with Eqs.~(\ref{eq:mom},\ref{eq:psiansatz}), while  the subscript Cal refers to that with Eqs.~(\ref{eq:psi0}-\ref{eq:mom}). The obtained parameters $\alpha, \beta$ and $c_i$  vary with $|\vect{k}_T|$, reflecting the change of shape in $x$ as $|\vect{k}_T|$ varies in $\psi_0$ and $\psi_1$ . Finally the full $\psi_0(x,\vect{k}_T^2)$ and $\psi_1(x,\vect{k}_T^2)$ can be obtained by interpolating the $\psi$'s at discretized $\vect{k}_T$'s.

\begin{figure}[tbp]
\centering\includegraphics[width=\columnwidth]{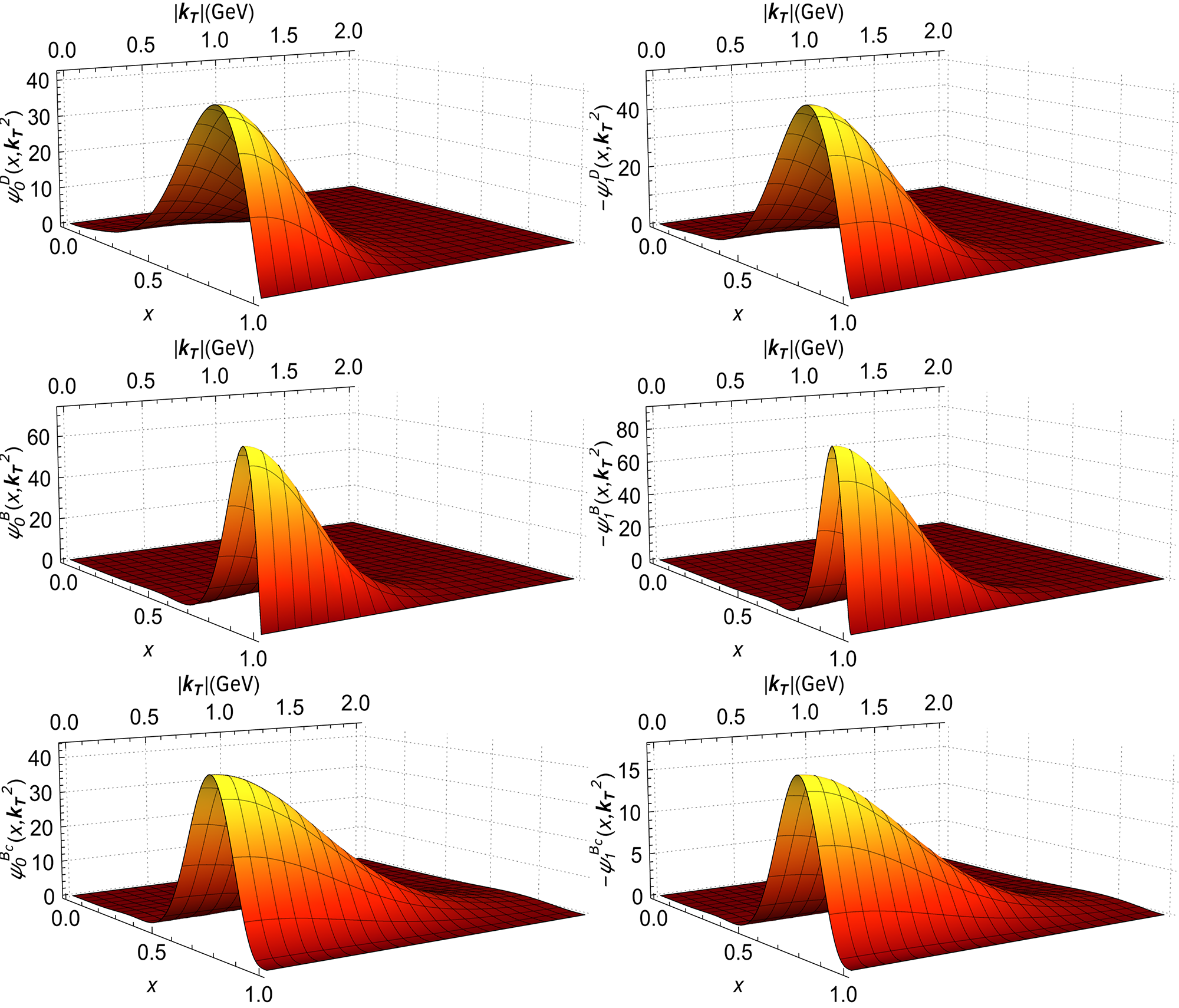}
\caption{The $\psi_0(x,\vect{k}_T^2)$ (left column) and $-\psi_1(x,\vect{k}_T^2)$ (right column) of $D$ meson (top row), $B$ meson (middle row) and $B_c$ meson (bottom row). }
\label{fig:psisq}
\end{figure}

We show the LF-LFWFs of $D$, $B$ and $B_c$ meson in Fig.~\ref{fig:psisq}. A characteristic feature is that these LFWFs are heavily tilted to larger $x$, meaning heavier quark tends to carry more light front momentum fraction of the parent hadron. As $\vect{k}_T$ goes larger, we find the LFWFs are less tilted, and tends to be more symmetric in $x$. This had also been found by BLFQ study in \cite{Tang:2019gvn}, showing a non-separable $x$ and $\vect{k}_T$ dependence in the LF-LFWFs. Such a behavior is partly due to that the light front kinematical energy of quark is $k^-=\frac{\vect{k}_T^2+m_q^2}{2k^+}$, hence a larger $\vect{k}_T$ helps reducing the  differences brought by current quark masses. Nevertheless, we remind that the narrow distribution in $x$ is a characteristic feature of LFWFs of a heavy meson, as found in $\eta_c$, $\eta_b$ and $J/\psi$ \cite{Shi:2021nvg}. A natural question to ask is then whether the heavy-light ($Q\bar{q}$) mesons LFWFs inherit any feature from light mesons. Our finding is that the $\vect{k}_T$-dependence of $Q\bar{q}$ LFWFs are actually closer to the $q\bar{q}$ mesons rather than $Q\bar{Q}$ ones. For instance, at $x \approx x_M$, with $x_M$ defined as the point where $|\psi_{i}(x,\vect{k}_T^2=0)|$ takes maximum value, the $\psi_{i}(x_M,\vect{k}_T^2)/\psi_{i}(x_M,\vect{k}_T^2=0)$ of $D$ meson is closer to that in pion rather than $\eta_c$. An analogous yet more detailed analysis will be addressed in section \ref{sec:3d}. 
%===============================================================================
%===============================================================================

%===============================================================================\end{document}
Generally speaking, all Fock-states light front wave functions of a meson should normalize to one, corresponding to the quark number sum rule.  In our case, the contribution of these LF-LFWF to the total normalization
\begin{align}
N_{\lambda,\lambda'}&=\int_0^1 dx \int \frac{d \vect{k}_T^2}{2(2 \pi)^3}  |\Phi_{\lambda,\lambda'}(x,\vect{k_T})|^2. \label{eq:norm}
\end{align}
are $(N_{\uparrow,\downarrow},N_{\uparrow,\uparrow})=(N_{\downarrow,\uparrow},N_{\downarrow,\downarrow})=(0.25,0.08)$, $(0.30,0.08)$, $(0.46,0.04)$ for $D$, $B$ and $B_c$ mesons respectively. We note that the $\sum_{\lambda,\lambda'}N_{\lambda,\lambda'}<1$ implies there are higher Fock-states within the heavy-light mesons. Within our approach, this includes higher Fock-states with gluons, as the gluons also enter the DS-BSEs as explicit degrees of freedom in Eq.~(\ref{eq.gluon}). However, the extraction of these higher Fock-state LFWFs from BS wave functions remains to be explored.

\begin{figure}[htbp]
\centering\includegraphics[width=\columnwidth]{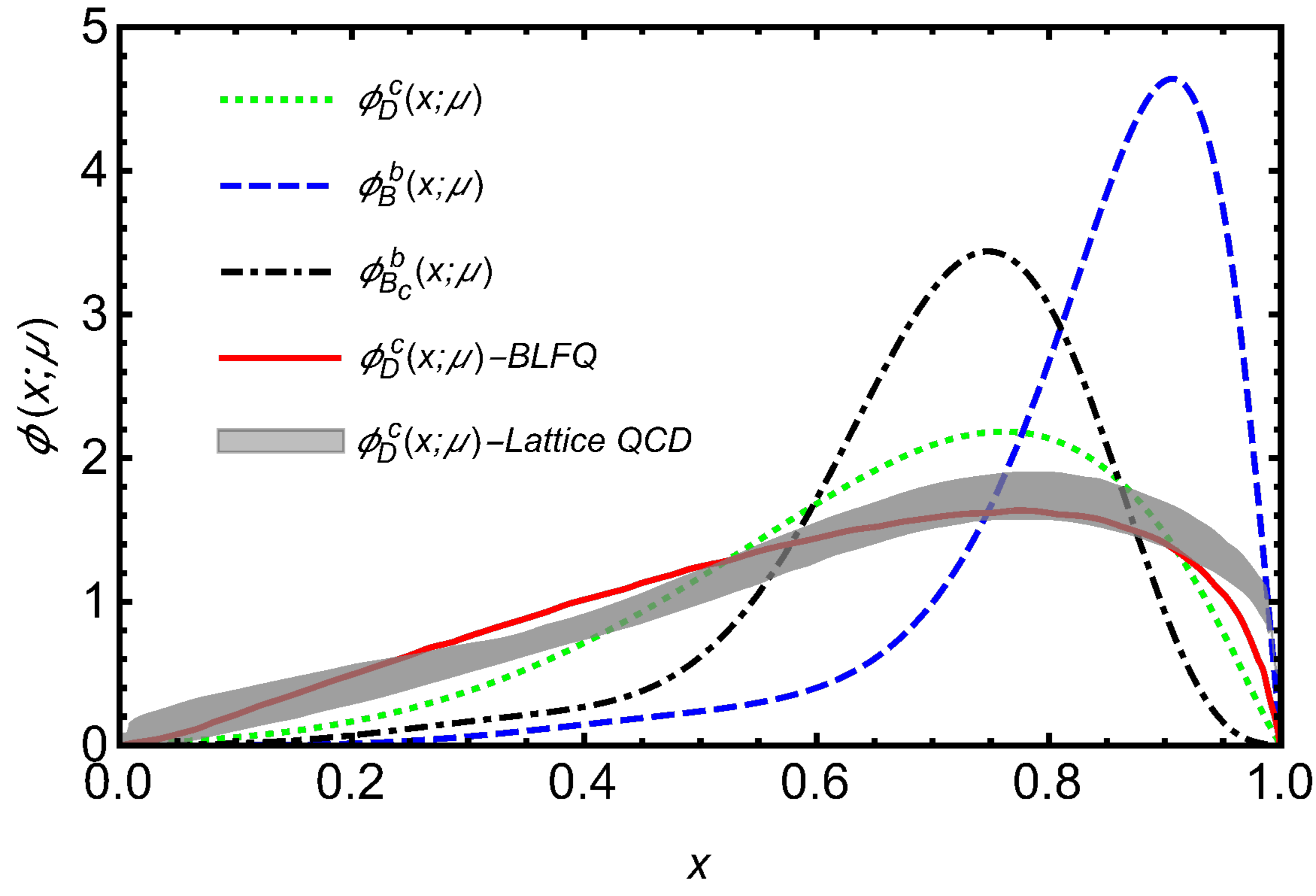} \caption{Parton distribution amplitudes (DAs) of $D$, $B$ and $B_c$ mesons. The lattice and BLFQ results are taken from \cite{Han:2024min} and \cite{Tang:2018myz} respectively.}
\label{fig:PDA}
\end{figure}

It is straightforward to look into the leading-twist  distribution amplitudes (DA) of the mesons, defined as the $\vect{k}_T$-integrated LFWF \cite{Lepage:1980fj}, i.e.,
\begin{align}
\label{eq:PDAdef}
\phi(x,Q)&\propto \int_{\vect{k}_T^2 \le Q^2}\frac{d^2 \vect{k}_T}{16 \pi^3}\  \psi_0(x,\vect{k}_T^2),
\end{align}
normalized by
\begin{align}
\int_0^1 dx\phi(x,Q)=1.
\end{align}
In Fig.~\ref{fig:PDA}, we show our results for the DAs of $D$ (green dotted), $B$ (blue dashed) and $B_c$ (black dot-dashed) mesons. The scale is set to be the sum of Euclidean constituent masses of meson's  (anti)quarks. Here the Euclidean constituent mass $M_f$ of a quark with flavor $f$ is defined as the mass that satisfies $M(p^2=M_f^2)=M_f$, with $M(p^2)\equiv B(p^2)/A(p^2)$ the mass function of dressed quark propagator. Note that this is not on-shell condition as we take the Euclidian space, and hence confinement remains respected. We therefore have $Q\approx (M_u+M_c)=1.7$ GeV for $D$ meson, and $Q\approx 4.6$ and $5.7$ GeV for $B$ and $B_c$ meson respectively. We find our results are close to earlier predictions from DSEs in \cite{Binosi:2018rht,Serna:2020txe}. This is encouraging as the interaction model and DA extraction techniques are a bit different in these papers. For instance, Ref.~\cite{Binosi:2018rht} took an extrapolation method in getting the DAs, and \cite{Serna:2020txe} used the Nakanishi-like representation to calculate the moments of DAs. Comparing with other calculations, we find our $D$ meson DA is narrower and taller than that from BLFQ (red solid), while the $B$ meson DAs are close \cite{Tang:2019gvn}. Very recently, lattice QCD reported $D$ meson quasi-DA (grey band) using large-momentum effective theory (LaMET) \cite{Han:2024min}, which is generally between BLFQ and our results. Note that the $D$ meson DA in \cite{Serna:2020txe} is a bit closer to lattice result, which can be viewed as a lower bound of $D$ meson DA based within DS-BSE approach. Meanwhile, Efremov–Radyushkin–Brodsky–Lepage (ERBL)  evolution \cite{Efremov:1978rn,Lepage:1979zb} would further bring down our DA curve as $Q$ increases.

\section{Three-dimensional Parton distribution functions of the heavy-light pseudoscalar mesons.}\label{sec:3d}

The three-dimensional parton distribution functions, such as genealized parton distribution functions and transverse momentum dependent parton distribution functions, have gained growing interest in the study of hadron structure as they encode rich information of parton motion, spatial distribution and spin-orbit correlation inside hadron. Although their experimental study is limited to very few stable hadrons such as nucleon and pion, out of theoretical interest here we utilize the GPDs and TMDs to investigate the 3D parton image of the heavy-light mesons.

The leading twist unpolarized quark GPD of a pseudoscalar meson in light-cone gauge is defined as
\begin{align}
H^q_M(x,\xi,t;\mu) &= \frac{1}{2}\int\frac{dz^-}{2\pi}\,e^{ixP^+z^-}  \nonumber \\
&\hs*{5mm}
\left< P+\tfrac{\Delta}{2}\left|\bar{\psi}^q(-\tfrac{z^-}{2})\,\gamma^+\,\psi^q(\tfrac{z^-}{2})\right|P-\tfrac{\Delta}{2}\right>. \label{eq:Hdef}
\end{align}
The $x$ is the parton's averaged light-cone momentum fraction and $\xi=-\frac{\Delta^+}{2 P^+}$ is the skewness. The $t = \Delta^2 = -\frac{4\xi^2 m_M^2+\vect{\Delta}_T^2}{1-\xi^2}$ is the squared four momentum transfer. The factorization scale $\mu$ in our case is taken to be the same as that of DA. Here we focus on the unpolarized GPD at zero skewness, i.e., the $H^q_M(x,0,t)$, as it is connected with the spatial distribution of quarks in hadron. The overlap representation of $H^q_M(x,0,t)$ in terms of LF-LFWFs reads ~\cite{Diehl:2003ny,Diehl:2000xz,Mezrag:2016hnp,Chouika:2016cmv}
\begin{align}
\label{eq:Hoverlapq}
H^q_M(x,0,t;\mu_0)&=\int \frac{d^2\vect{k}_T}{(2\pi)^3}\big[ \psi_0^*(x,\hat{\vect{k}}_T)\,\psi_0(x,\tilde{\vect{k}}_T)\nonumber \\
&\hspace{12mm}
+\hat{\vect{k}}_T \cdot \tilde{\vect{k}}_T\,\psi_1^*(x,\hat{\vect{k}}_T)\,\psi_1(x,\tilde{\vect{k}}_T) \big],
\end{align}
with $\hat{\vect{k}}_T=\vect{k}_T+(1-x)\frac{\vect{\Delta}_T}{2}$ and $\tilde{\vect{k}}_T=\vect{k}_T-(1-x)\frac{\vect{\Delta}_T}{2}$. For the antiquark with flavor $h$, 

\begin{align}
\label{eq:Hoverlapqbar}
H^{h}_M(x,0,t;\mu_0)&=-\int \frac{d^2\vect{k}_T}{(2\pi)^3}\big[ \psi_0^*(x',\hat{\vect{k}}'_T)\,\psi_0(x',\tilde{\vect{k}}'_T)\nonumber \\
&\hspace{12mm}
+\hat{\vect{k}}'_T \cdot \tilde{\vect{k}}'_T\,\psi_1^*(x',\hat{\vect{k}}'_T)\,\psi_1(x',\tilde{\vect{k}}'_T) \big],
\end{align}
with $x'=1+x$, $\hat{\vect{k}}_T=\vect{k}_T+(1-x')\frac{\vect{\Delta}_T}{2}$ and $\tilde{\vect{k}}_T=\vect{k}_T-(1-x')\frac{\vect{\Delta}_T}{2}$. Note that Eqs.~(\ref{eq:Hoverlapq},\ref{eq:Hoverlapqbar}) impose the leading Fock-state truncation, hence from now on we rescale the LF-LFWFs of heavy-light mesons by an overall factor so that they would saturate the normalization condition, i.e., $\sum_{\lambda,\lambda'} N_{\lambda,\lambda'}=1$. In terms of GPD, this ensures $\int_0^1 dx H^q_{M}(x,0,0;\mu_0)=\int_0^1 dx f^q_{M}(x;\mu_0)=1$, with $f(x;\mu_0)$ the collinear unpolarized parton distribution functions. This means we rescale the LF-LFWFs of $D$, $B$ and $B_c$ mesons by multiplying $1.23$, $1.15$ and $1.0$ respectively, according to values reported below Eq.~(\ref{eq:norm}).

%===============================================================================

\begin{figure}[tbp]
\centering\includegraphics[width=0.98\columnwidth]{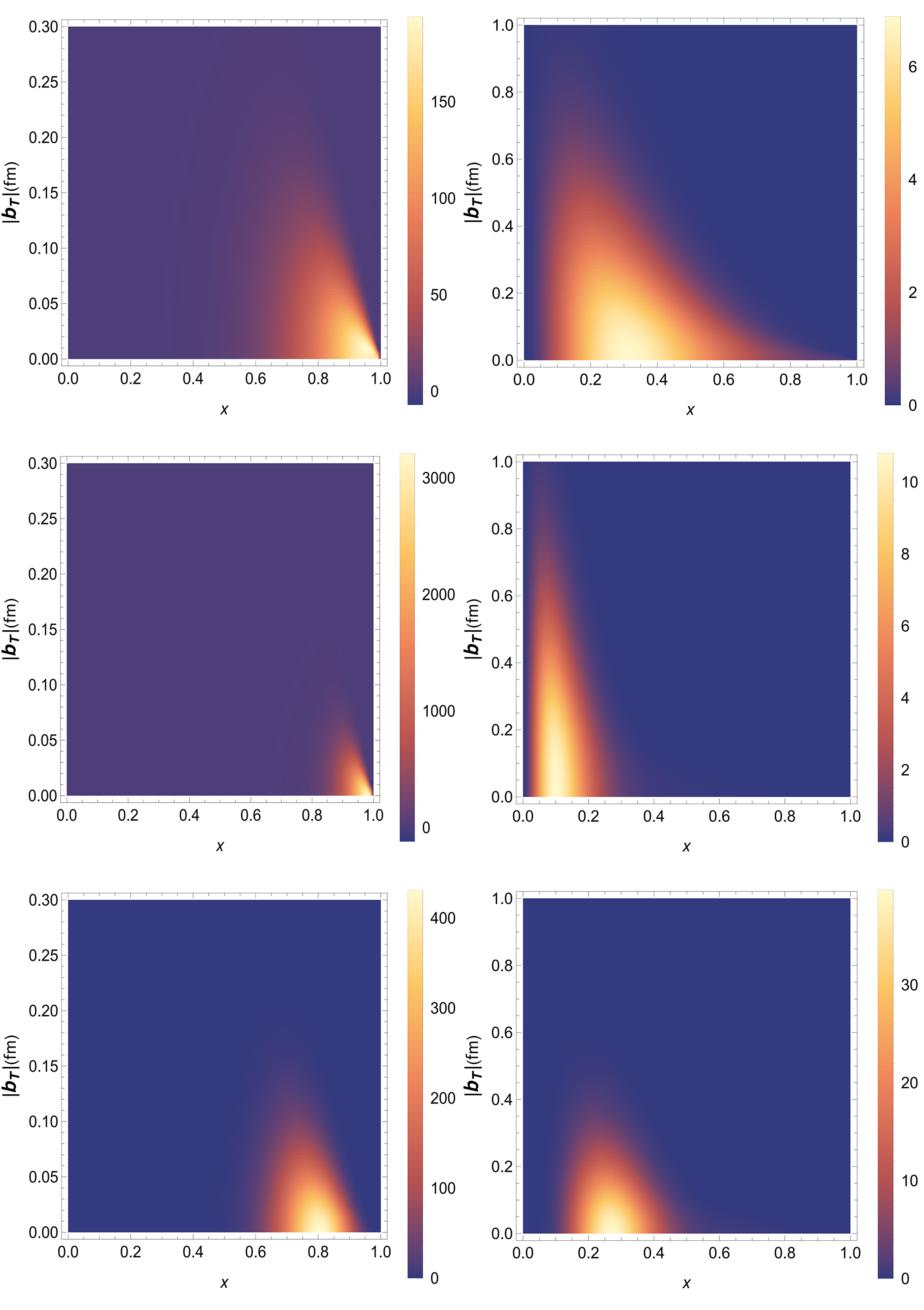} 
\caption{The IPD GPD $\rho(x,\vect{b}_T^2)$ for $D$  (top row), $B$  (middle row) and $B_c$ (bottom row) mesons. The left column is for heavier quark and right column for the lighter one.}

\label{fig:rhos}
\end{figure}

\begin{figure}[tbp]
\centering\includegraphics[width=0.98\columnwidth]{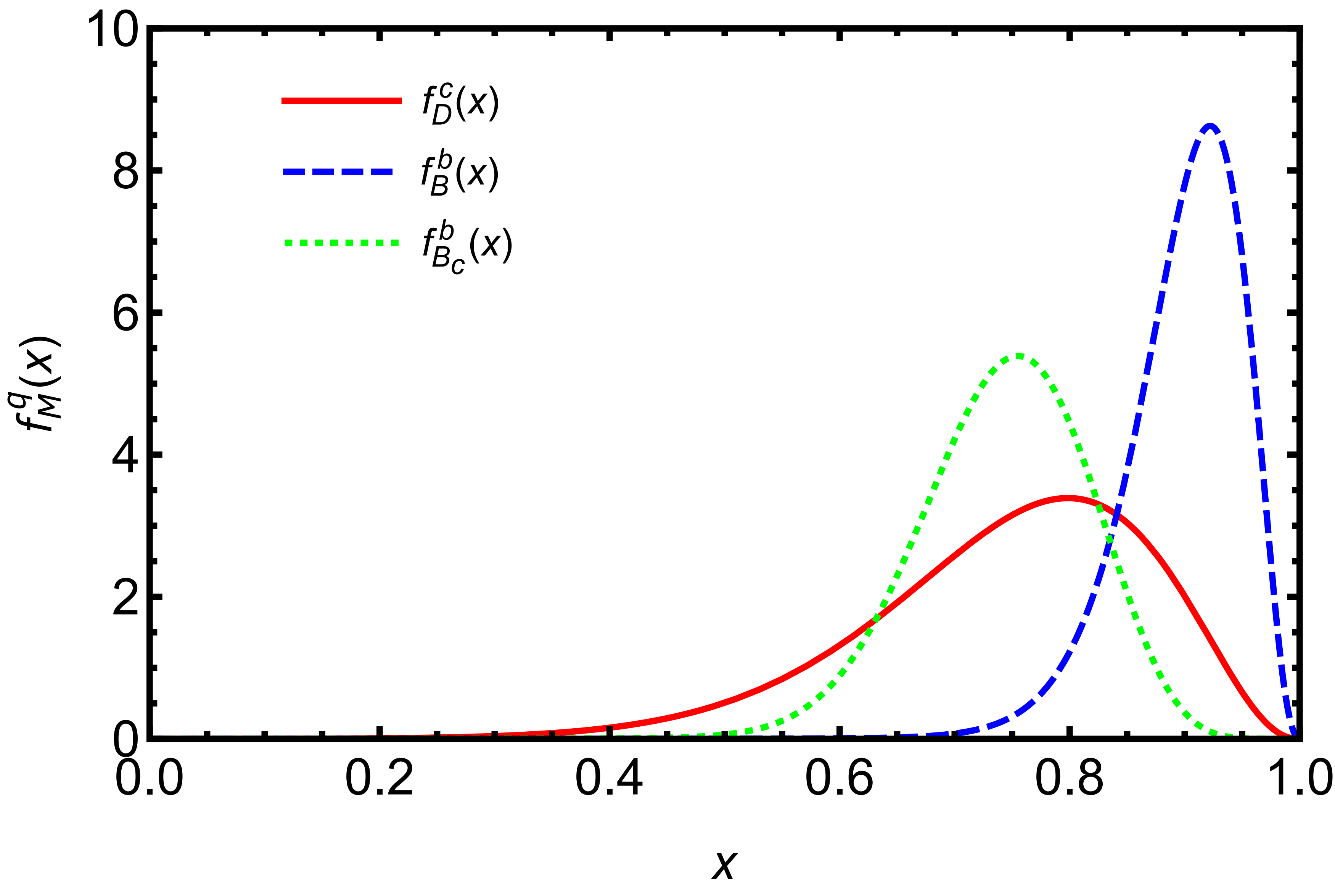} 
\caption{\looseness=-1 The unpolarized collinear parton distribution function of heavier quark in $D$, $B$ and $B_c$ mesons.}
\label{fig:PDF}
\end{figure}

Performing a two-dimensional Fourier transform of $H_M^q(x,0,\Delta_T^2)$, one gets the impact parameter dependent (IPD) GPD
\begin{align}
\rho_M^q(x,\vect{b}_T^2) = 
\int \frac{d^2 \vect{\Delta}_T}{(2\pi)^2}\,H_M^q(x,0,-\vect{\Delta^2_T})\,e^{i \vect{b}_T \cdot \vect{\Delta}_T }, \label{eq:ipdfourier}
\end{align}
which characterizes the probability density of partons on the transverse plane with  the light-cone momentum fraction $x$ and the impact parameter $\vect{b}_T$ \cite{Burkardt:2002hr}. We show our results in Fig.~\ref{fig:rhos}, with the rows  devoted to the $D$, $B$ and $B_c$ mesons from top to bottom, and columns for heavier and lighter quark from left to right. Note that these IPD GPDs are associated with their hadronic scales. 
Integrating over the $\vect{b}_T$, one gets the collinear parton distribution functions, which are shown in Fig.~\ref{fig:PDF}. We only show the curves for heavier quark distribution $f_f(x)$, and the lighter quark distribution can be obtained as $f_{\bar{h}}(x)=f_f(1-x)$ due to momentum conservation. Apparently, in the heavy-light mesons the heavier quark carries most light-front momentum, producing a prominent peak in the large $x$ region.

\begin{figure}[htbp]
\centering\includegraphics[width=\columnwidth]{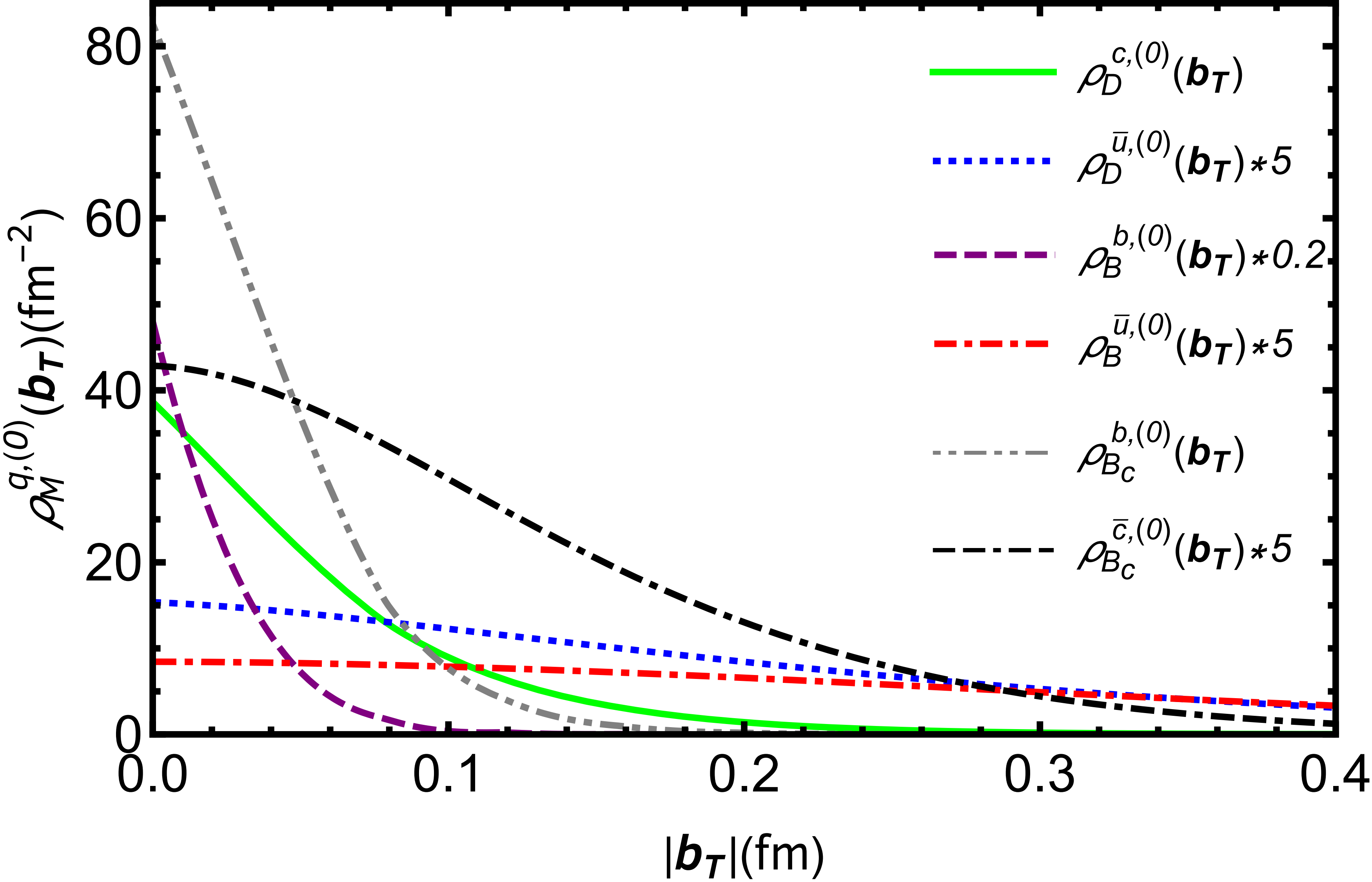} \\
\caption{\looseness=-1 The spatial quark distribution (see Eq.~(\ref{eq:rho0})) in the transverse plane within $D$, $B$ and $B_c$ mesons at hadronic scales.}
\label{fig:rho0}
\end{figure}

On the other hand, one can obtain the spatial parton distributions in the transverse plane by integrating over $x$ in the IPD GPD, i.e.,
\begin{align}\label{eq:rho0}
 \rho_M^{q,(0)}(\vect{b_T})=\int_0^1 dx \rho_M^q(x,\vect{b_T}^2).
\end{align}
 The results are shown in Fig.~\ref{fig:rho0}. Generally speaking, the heavier quarks are much more centered in the considered flavor-asymmetric mesons. Note that, if the curves are viewed as the valence quark distribution, i.e., quark distribution minus anti-quark distribution, then they are actually scale independent due to the evolution properties of IPD GPDs. Defining their mean squared impact parameter $\langle \vect{b}_T^2 \rangle^q= \int d \vect{b}_T^2 \vect{b}_T^2 \rho^{q,(0)}(\vect{b}_T^2)$ as a measure for their spatial extension, we find $\langle \vect{b}_T^2 \rangle^c_{D}=(0.13\ {\rm fm})^2$, $\langle \vect{b}_T^2 \rangle^{\bar{u}}_{D}=(0.40\ {\rm fm})^2$, $\langle \vect{b}_T^2 \rangle^b_{B}=(0.056\ {\rm fm})^2$, $\langle \vect{b}_T^2 \rangle^{\bar{u}}_{B}=(0.47\ {\rm fm})^2$ and $\langle \vect{b}_T^2 \rangle^b_{B_c}=(0.077\ {\rm fm})^2$, $\langle \vect{b}_T^2 \rangle^{\bar{c}}_{B_c}=(0.22\ {\rm fm})^2$, in comparison with our results in pion and heavy quarkonium $\langle \vect{b}_T^2 \rangle^u_{\pi}=(0.33\ {\rm fm})^2$ \cite{Shi:2020pqe}, $\langle \vect{b}_T^2 \rangle^c_{\eta_c}=(0.16\ {\rm fm})^2$ and $\langle \vect{b}_T^2 \rangle^b_{\eta_b}=(0.092\ {\rm fm})^2$ \cite{Shi:2021nvg}. We find the $\bar{u}-$quark in $D$ meson is more spread out than that in pion, and the $c-$quark in $D$ meson is more centered than that in $\eta_c$ meson. Same picture is also found in $B$ and $B_c$ mesons. This drastic distribution difference between the heavy and light quarks indicates that the light quarks could play a more important role in determining the overall charge and/or mass distribution within flavor-asymmetric mesons.

% u,c,t:+2/3 
% d,s,b:-1/3 

With $\langle \vect{b}_T^2 \rangle^q_M$, we can obtain the charge radius of meson in the light-cone frame. Here we only consider the charged meson. For instance, the light-cone charge radius of $D^+$ is $\langle r^2_{c,LC} \rangle_{D^+}=e_c \langle \vect{b}_T^2 \rangle^c_{D^+}+e_{\bar{d}} \langle \vect{b}_T^2 \rangle^{\bar{d}}_{D^+}=(0.25\ {\rm fm})^2$. Analogously, we obtain $\langle r^2_{c,LC} \rangle_{B^-}=(0.38\ {\rm fm})^2$ ($b\bar{u}$), $\langle r^2_{c,LC} \rangle_{B_c^-}=-(0.19 \ {\rm fm})^2$ ($d\bar{b}$). The negative sign can be considered as negative charge distribution. There are two points to be noted. First, this radius is well defined in the sense of characterizing charge density distribution in the light-cone frame, and should be distinguished from the conventional charge radius defined in the Breit frame, i.e., $\langle r^2\rangle=-6\frac{dF(Q^2)}{dQ^2}$ with $F(Q^2)$ the eletro-magnetic form factor. Their differences had been discussed with detail in \cite{Miller:2018ybm}. Here we refrain from giving the $F(Q^2)$ because \cite{Xu:2024fun} had given calculation based on full covariant DS-BSEs, and meanwhile a calculation based on GPD approach would receive a nontrivial zero mode contribution proportional to $\delta(x)$ that complicates the problem (see appendix of \cite{Shi:2020pqe}). Secondly, in heavy-light mesons the light-cone charge radii are strongly affected by the light quarks, e.g., the $B^-$, as the  $\langle \vect{b}_T^2 \rangle$ of light quarks are much larger than heavy quarks.  

\begin{figure}[htbp]
\centering\includegraphics[width=\columnwidth]{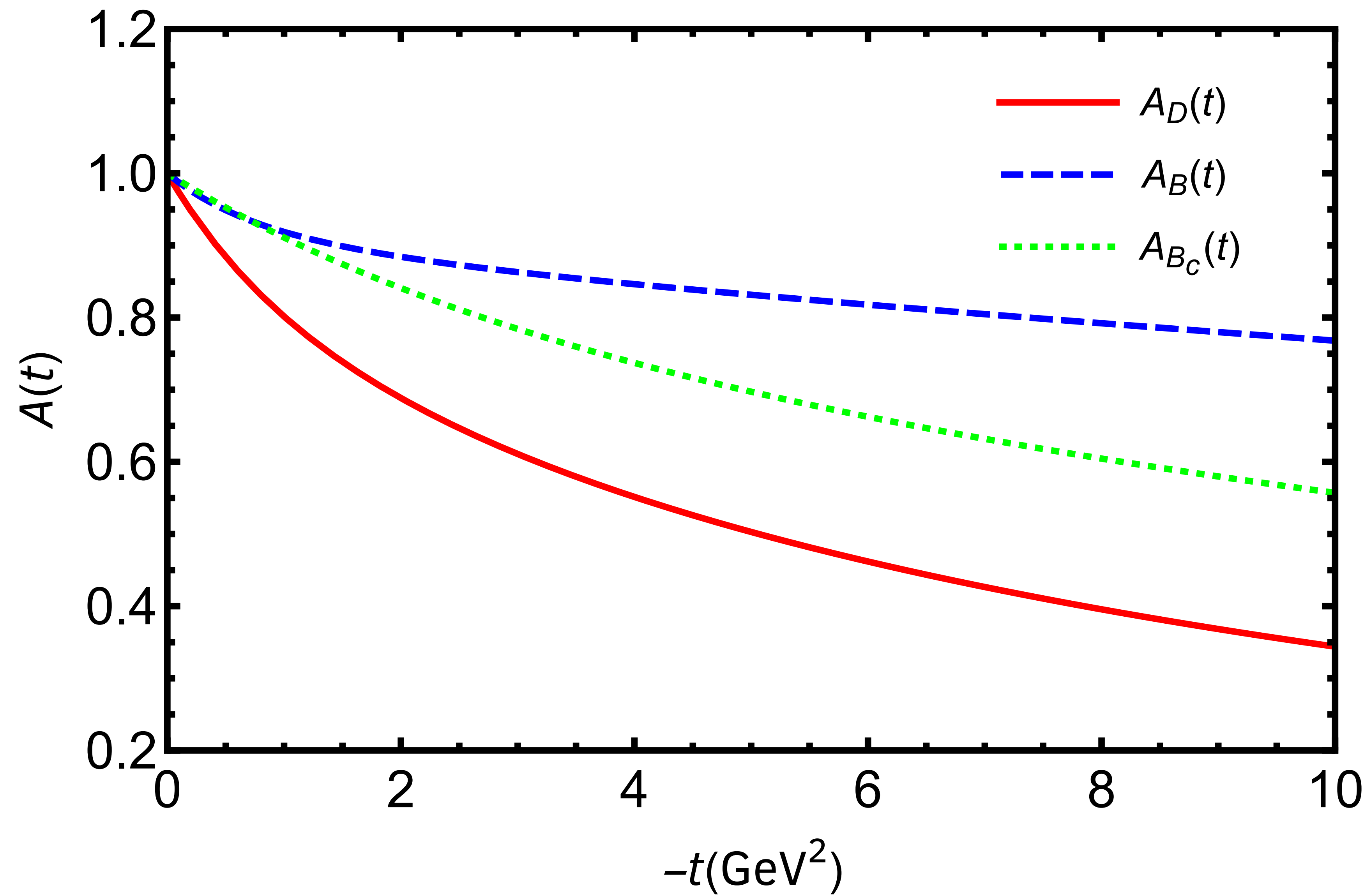} 
\caption{\looseness=-1 The gravitational form factor $A(t)$ (see definition in Eq.~(\ref{eq:EMT})) of $D$, $B$ and $B_c$ mesons.
\label{fig:GFF}}
\end{figure}

\begin{figure}[htbp]
\centering\includegraphics[width=\columnwidth]{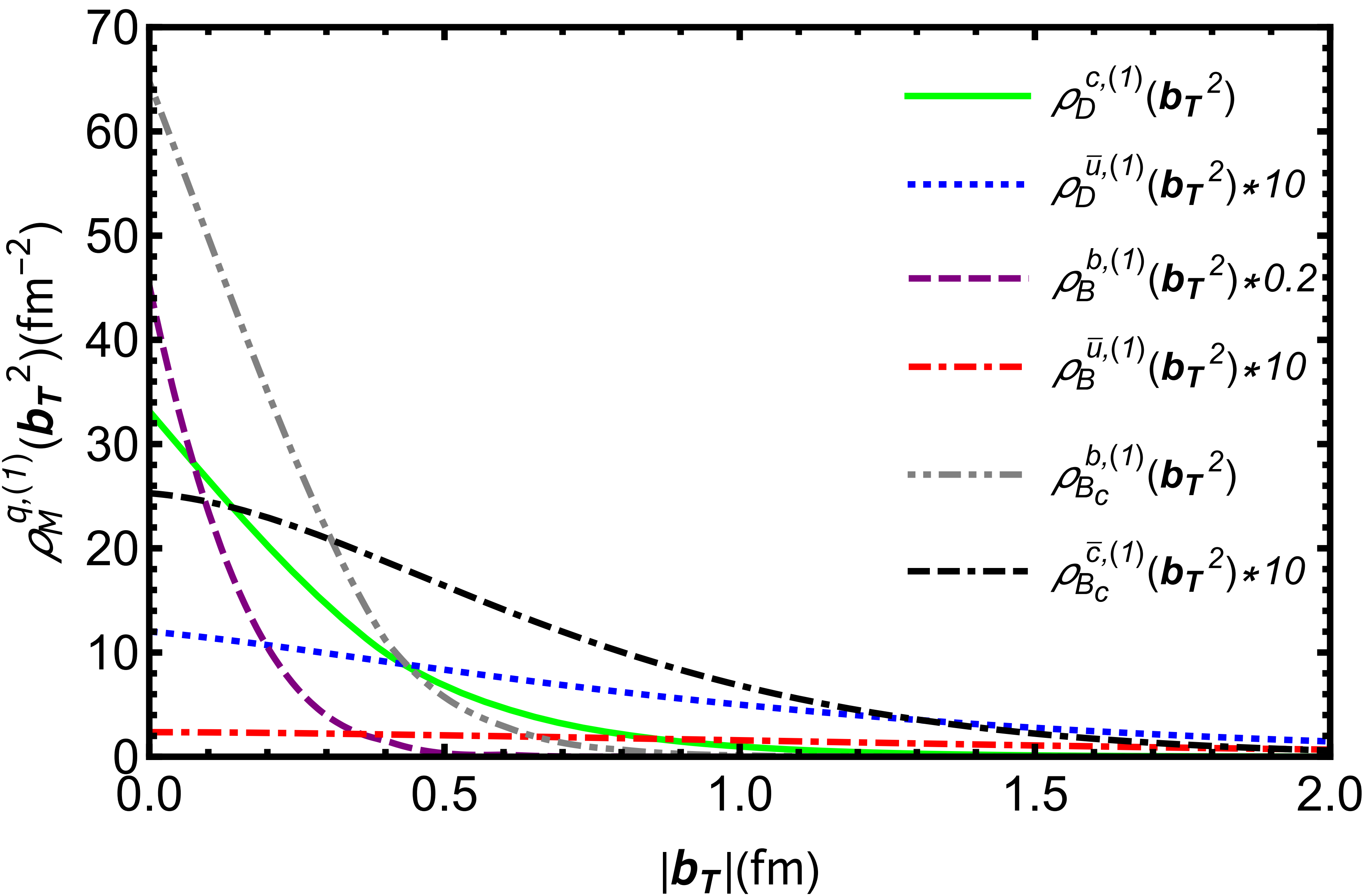} 
\caption{\looseness=-1 The spatial distribution of light-cone momentum fraction density $\rho_M^{q,(1)}(\vect{b}_T^2)$ (see definition in Eq.~(\ref{eq:rho1})) in the transverse plane within $D$, $B$ and $B_c$ mesons at hadronic scales.
\label{fig:rho1}}
\end{figure}

We can also study the mass distribution within heavy-light mesons with gravitational form factor (GFF), which is related to the unpolarized GPD. The GFFs of pesudoscalar meson are defined as coefficient functions in the general decomposition of the matrix element of the energy-momentum tensor (EMT) between spin-0 states~\cite{Donoghue:1991qv,Polyakov:2018zvc,Freese:2021mzg}
\begin{align}
\langle M(p')|T^{\mu\nu}(0)|M(p)\rangle&=\frac{1}{2}[P^\mu P^\nu A(t) +(g^{\mu \nu}q^2-q^\mu q^\nu)C(t)].
\label{eq:EMT}
\end{align}
with $P=p+p'$, $q=p'-p$ and $t=q^2$. The $A(t)$ can be decomposed into contributions from different partons, i.e., $A(t) = \sum_a A_{2,0}^a(t;\mu)$, with $a$ running through all quarks and gluons. In our case, the gluon is absent due to the imposed leading Fock-state approximation, hence $A_{2,0}^g=0$. The quark GFF can be obtained with 
\begin{align}
\int_{-1}^1dx\, x\, H^q(x,0,t;\mu_0) &= A^q_{2,0}(t;\mu_0).
\label{eq:theta2}
\end{align}
Summing up the valence quark and antiquark contributions, we obtain the $A(t)$'s of flavor-asymmetric mesons as displayed in Fig.~\ref{fig:GFF}, which is scale independent. From the GFF $A(Q^2)$, we can determine the light-cone mass radius $\langle r_{E,{\rm LC}}^2 \rangle$, which is defined as the mean value of $\vect{r}_\perp^2$ weighted by the EMT in the light-cone frame, with \cite{Freese:2019bhb} 
\begin{align}
\langle r_{E,{\rm LC}}^2 \rangle = -4\,\lf.\frac{\partial\, A(Q^2)}{\partial Q^2} \rg|_{Q^2=0},
\end{align}
with $Q^2=-t$. From Fig.~\ref{fig:GFF} we extract $\langle r_{E,{\rm LC}}^2 \rangle_D=(0.21\ {\rm fm})^2$, $\langle r_{E,{\rm LC}}^2 \rangle_B=(0.14 \ {\rm fm})^2$ and $\langle r_{E,{\rm LC}}^2 \rangle_{B_c}=(0.13\ {\rm fm})^2$, as compared to $\langle r_{E,{\rm LC}}^2 \rangle_{\pi}\approx (0.27 \ {\rm fm})^2$, $\langle r_{E,{\rm LC}}^2 \rangle_{\eta_c}=(0.15 \ {\rm fm})^2$ and $\langle r_{E,{\rm LC}}^2 \rangle_{\eta_b}=(0.089\ {\rm fm})^2$ \cite{Kumano:2017lhr,Freese:2019bhb,Shi:2021nvg}. This is an outcome of the competition  between the heavier quark, which carries most of the light-cone energy and is concentrated at the center, and the surrounding lighter quark cloud, which contains less energy but is spatially more extended. To see that more clearly, we look into the first moment of IPD GPD, i.e., 
\begin{align} \label{eq:rho1}
\rho_M^{q,(1)}(\vect{b}_T^2)=\int_0^1 dx x \rho_M^q(x,\vect{b_T}^2) 
\end{align}
 It is easy to see that it is actually the Fourier transform of $A^q_{2,0}(t)$ with respect to $\vect{\Delta}_T$, ($t=-\vect{\Delta}_T^2$), by inspecting Eqs.~(\ref{eq:rho1},\ref{eq:ipdfourier},\ref{eq:theta2}). Its physical picture is also obvious: it indicates the density of light-cone momentum fraction distribution in the transverse plane within a meson. We show our results in Fig.~\ref{fig:rho1}. Some of the curves are also rescaled for better display. We note that the light-cone energy radius $\langle r_{E,{\rm LC}}^2 \rangle$ is equivalent to $\sum_q \int d\vect{b}_T^2 \vect{b}_T^2 \rho_M^{q,(1)}(\vect{b}_T^2)$. One can see from Fig.~\ref{fig:rho1} how the mass radius are affected by the light and heavy quarks.

%===============================================================================

\begin{figure}[htbp]
\centering\includegraphics[width=\columnwidth]{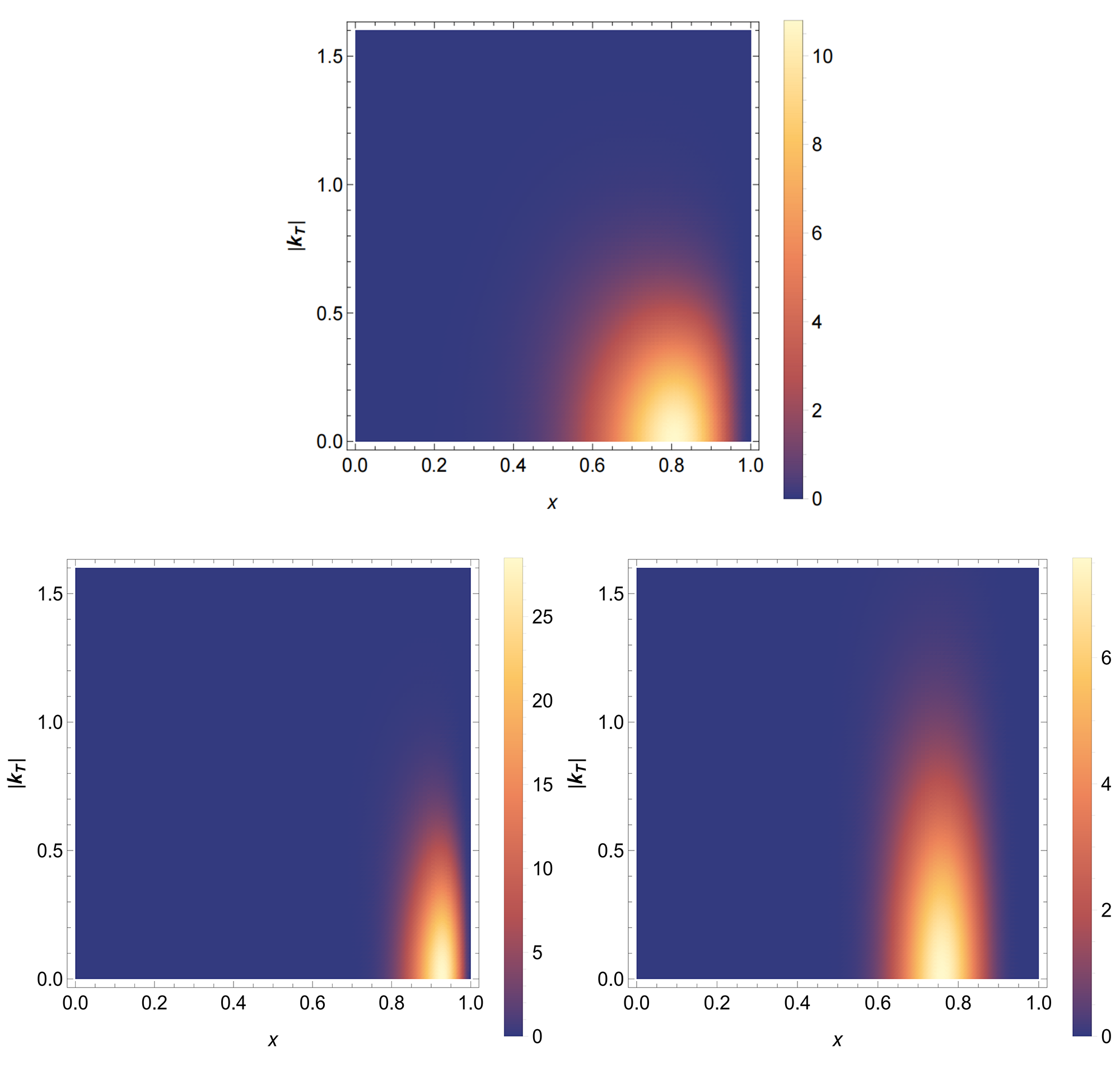} \\
\caption{
Density plots of the unpolarized TMD PDF $f_1^q(x,\vect{k}_T^2)$ of $D$ (top row), $B$ (lower left) and $B_c$ (lower right) mesons.}
\label{fig:TMD}
\end{figure}

We finally investigate the transverse momentum distribution of quarks within flavor-asymmetric mesons. This can be revealed by the unpolarized leading twist-even TMD, which is defined as
\begin{align} 
f_{1}(x,\vect{k}_T^2)&=\int\frac{d \xi^-d^2\vect{\xi}_T}{(2 \pi)^3}\ e^{i(\xi^-k^+-\vect{\xi}_T\cdot \vect{k}_T)}\nonumber\\
&\hspace{28mm}
\langle P|\bar{\psi}(0)\gamma^+\psi(\xi^-,\vect{\xi}_T)|P\rangle,
\end{align}
with the gauge link approximated to be unity. In the leading Fock-state truncation, its overlap representation in terms of LFWFs reads~\cite{Pasquini:2014ppa}
\begin{align}
f^q_1(x,\vect{k}_T^2) = \frac{1}{(2 \pi)^3} 
\left[\lf|\psi_0(x,\vect{k}_T^2)\rg|^2 + \vect{k}_T^2 \lf|\psi_1(x,\vect{k}_T^2)\rg|^2\right],
\label{eq:tmd}
\end{align}
which has the probabilistic interpretation of one quark carrying longitudinal momentum fraction $x$ and transverse momentum $\vect{k}_T$. Consequently the antiquark carries longitudinal momentum fraction $1-x$ and transverse momentum $-\vect{k}_T$. We show the density plots of $D$ (upper left), $B$ (upper right) and $B_c$ (lower) TMDs in Fig.~\ref{fig:TMD}. These are generally narrow and skewed distributions in $x$. On the other hand, the mean transverse momenta of the valence quarks $\langle |\vect{k_T}| \rangle=\int dx d^2 \vect{k}_T f_1^q(x,\vect{k}_T^2) |\vect{k}_T|$ are found to be $\langle |\vect{k_T}| \rangle_{D}=0.43$ GeV, $\langle |\vect{k_T}| \rangle_{B}=0.42$ GeV and $\langle |\vect{k_T}| \rangle_{B_c}=0.65$ GeV, as compared to $\langle |\vect{k_T}| \rangle_{\pi}=0.39$ GeV, $\langle |\vect{k_T}| \rangle_{\eta_c}=0.65$ GeV and $\langle |\vect{k_T}| \rangle_{\eta_b}=1.0$ GeV at hadronic scales. As pointed out in Sec.~\ref{sec:bs2lf}, we find the transverse momentum distribution in $Q\bar{q}$ system is closer to $q\bar{q}$ rather than $Q\bar{Q}$.

%===============================================================================
\begin{figure}[htbp]
\centering\includegraphics[width=\columnwidth]{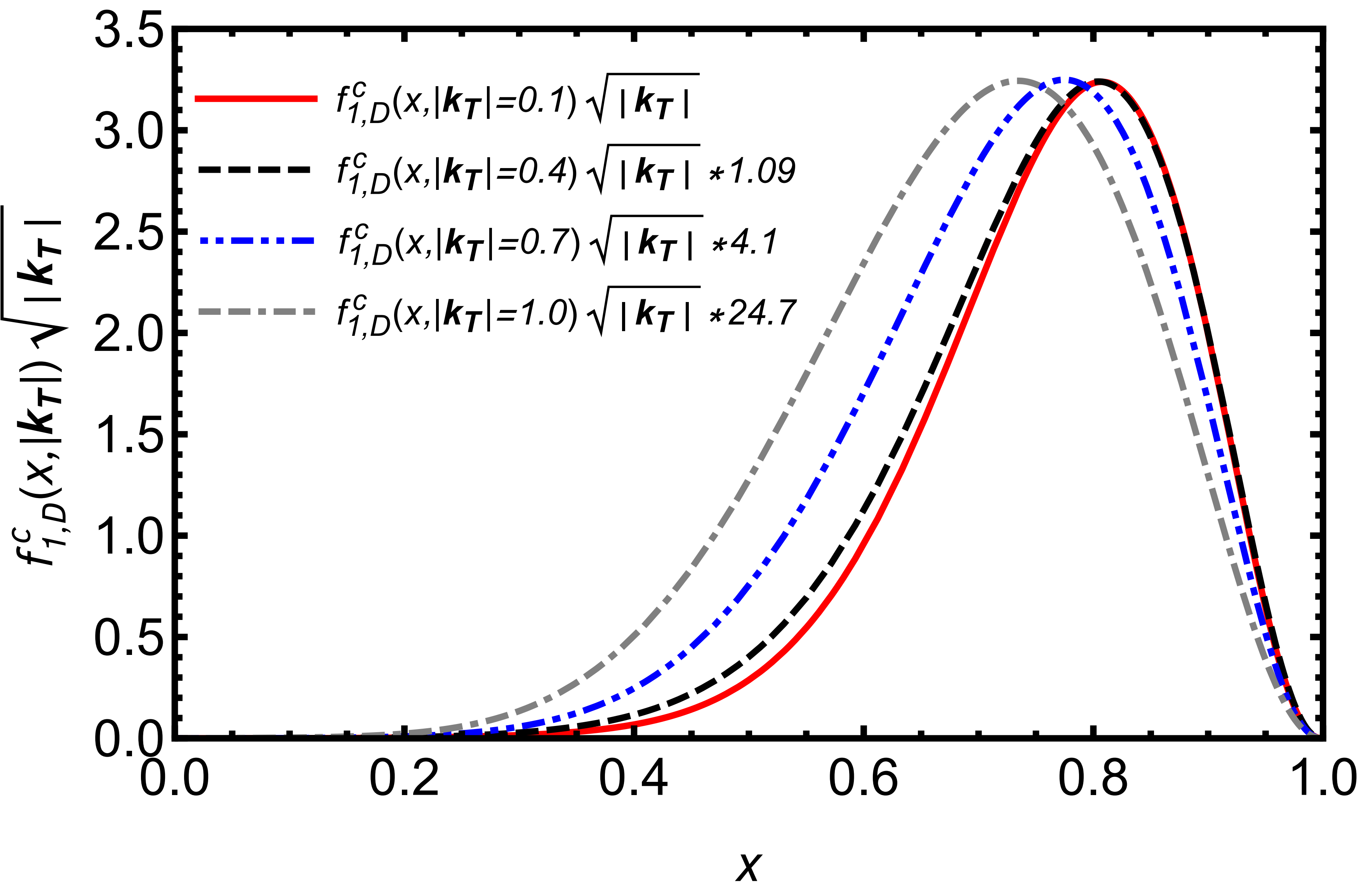} \\
\caption{
The unpolarized TMD PDF of charm quark in $D$ meson $f_{1,D}^c(x,\vect{k}_T^2)$ at different $|\vect{k}_T|$'s.}
\label{fig:TMD2}
\end{figure}

 Finally, we show the evolution of $x-$dependence in TMDs as $\vect{k}_T$ increases in Fig.~\ref{fig:TMD2}. Here we display $\sqrt{|\vect{k}_T|} f_1(x,\vect{k}_T^2)$. The factor $\sqrt{|\vect{k}_T|}$ provides a guidance on one TMD's contribution to physical observables. It comes from the integral measure $\int d\vect{k}_T^2=\int d|\vect{k}_T||\vect{k}_T|$, which shows up in the normalization condition $\int d\vect{k}_T^2 \int dxf_1(x,\vect{k}_T^2)=1$, as well as in the TMD convolutions $\int d\vect{k}_T^2 f_1(x,\vect{k}_T^2)\otimes {\cal F}(x,\vect{k}_T^2)$  for scattering structure functions with ${\cal F}$ denoting TMD PDF or TMD fragmentation functions. One can see that as $|\vect{k}_T|$ increases, the $x-$distribution becomes more symmetric, similar to the LF-LFWFs as mentioned in Sec.~\ref{sec:bs2lf}. Note that the curves are rescaled to same magnitude, with the rescaling factor given in the plot, among them $|\vect{k}_T|=0.4$ GeV is chosen to be close to the typical transverse momentum $\langle |\vect{k_T}| \rangle_{D}=0.43$ GeV. One can see the curves are similar when $|\vect{k}_T|<\langle |\vect{k_T}| \rangle_{D}$, while are deformed as $|\vect{k_T}|$ increases.

\section{CONCLUSION\label{sec:con}}
We study the leading Fock-state light front wave functions of heavy flavor asymmetric pseudoscalar mesons, e.g., $D$, $B$ and $B_c$, which are extracted from their covariant Bethe-Salpeter wave functions. For a heavy-light system $Q\bar{q}$, the LF-LFWFs inherit features from both $Q\bar{Q}$ and $q\bar{q}$ systems, e.g., they are narrow in $x$ which is closer to $Q\bar{Q}$, while their $|\vect{k}|_T$-distribution is closer to that in $q\bar{q}$,  exhibiting a duality embodying characteristics from both light mesons and heavy quarkonium. We then obtain the DAs and they are comparable with existing calculations, either earlier DS-BSEs study using different model setup and calculation techniques or BLFQ and lattice computation results.

We then impose the leading Fock-state truncation and study the hadron tomography of $D$, $B$ and $B_c$ mesons with unpolarized GPD and TMD. The impact parameter dependent GPD is utilized to study the spatial distribution of quark and antiquark. It is found that in flavor-asymmetric mesons $Q\bar{q}$, the heavier quarks' distributions are significantly more centered than those in $Q\bar{Q}$, while the light quarks are even more spread out than those in $q\bar{q}$, e.g., $\langle \vect{b}_T^2 \rangle^c_{D}=(0.13\ {\rm fm})^2$, $\langle \vect{b}_T^2 \rangle^{\bar{u}}_{D}=(0.40\ {\rm fm})^2$ as compared to  $\langle \vect{b}_T^2 \rangle^c_{\eta_c}=(0.16\ {\rm fm})^2$ and $\langle \vect{b}_T^2 \rangle^u_{\pi}=(0.33\ {\rm fm})^2$. The light-cone charge and energy radii that characterize the charge and energy spatial distributions are also given, with the former larger than the latter. There is also an extreme case as the $B^-$ meson, where the $\bar{u}$ quark is rather spread out, causing a light-cone charge radius much larger than its energy radius, e.g., $\langle r_{E,{\rm LC}}^2 \rangle_B=(0.14 \ {\rm fm})^2< \langle r^2_{c,LC} \rangle_{B^-}=(0.38\ {\rm fm})^2$ ($b\bar{u}$), as most energy is carried by the $b$ quark that sit in the center. Finally, regarding the transverse motion within the mesons, we find at hadronic scales, both heavier and lighter quarks tend to carry transverse momentum that are close to light quarks in light mesons. Meanwhile, the non-separable $x$ and $\vect{k}_T$ dependence is also observed at relatively large $\vect{k}_T$. 

This work therefore delivers a comprehensive study on the parton structure of heavy flavor-asymmetric mesons, and reveal a novel parton picture that is a joint outcome of light and heavy sectors of quantum chromodynamics.

\begin{acknowledgments}
We thank Shunzo Kumano and Yin-Zhen Xu for helpful discussions.  This work was supported by the Fundamental Research Funds for the Central Universities (under Grant No. 1227050553), Nanjing Overseas Scholars Innovation Project Selective Financial Support, Taishan Scholars Program and Shandong Excellent Young Scientists Fund Program (Overseas) (No. 2023HWYQ-106).
\end{acknowledgments}

%===============================================================================

\bibliography{QqP}

\end{document}